\documentclass[runningheads]{llncs}
\usepackage{graphicx}
\usepackage[breaklinks=true,hidelinks=true]{hyperref}
\usepackage{CJKutf8} 
\usepackage[russian,english]{babel}
\usepackage{enumerate}

\begin{document}

\title{Challenges of Linking Organizational Information in Open Government Data to Knowledge Graphs}
\titlerunning{Challenges of Linking Organizational Information}

\author{Jan Portisch\inst{1}\orcidID{0000-0001-5420-0663} \and
Omaima Fallatah\inst{2}\orcidID{0000-0002-5466-9119} \and
Sebastian Neumaier\inst{3}\orcidID{0000-0002-9804-4882} \and
Mohamad Yaser Jaradeh\inst{5}\orcidID{0000-0001-8777-2780} \and
Axel Polleres\inst{3,4}\orcidID{0000-0001-5670-1146}}

\authorrunning{J. Portisch et al.}

\institute{
Data and Web Science Group, University of Mannheim, Germany\\
    \email{jan@informatik.uni-mannheim.de} \and
Information School, The University of Sheffield, Sheffield, UK\\
   \email{oafallatah1@sheffield.ac.uk} \and
Vienna University of Economics and Business, Austria\\
    \email{\{sebastian.neumaier, axel.polleres\}@wu.ac.at} \and
Complexity Science Hub Vienna, Austria \and
	L3S Research Center, Leibniz University Hannover, Germany\\
    \email{jaradeh@l3s.de}
}
\maketitle
\begin{abstract}
Open Government Data (OGD) is being published by various public administration organizations around the globe. 
Within the metadata of OGD data catalogs, the publishing organizations (1) are not uniquely and unambiguously identifiable and, even worse, (2) change over time, by public administration units being merged or restructured. In order to enable fine-grained analyzes or searches on
Open Government Data on the level of publishing organizations, linking those from OGD portals to publicly available knowledge graphs (KGs) such as \textit{Wikidata} and \textit{DBpedia} seems like an obvious solution. Still, as we show in this position paper, organization linking faces significant challenges, both in terms of available (portal) metadata and KGs in terms of data quality and completeness.
We herein specifically highlight five main challenges, namely regarding (1) temporal changes in organizations and in the portal metadata, (2) lack of a base ontology for describing organizational structures and changes in public knowledge graphs, (3) metadata and KG data quality, (4) multilinguality, and (5) disambiguating public sector organizations.
Based on available OGD portal metadata from the \textit{Open Data Portal Watch}, we provide an in-depth analysis of these issues, make suggestions for concrete starting points on how to tackle them along with a call to the community to jointly work on these open challenges. 

.
\end{abstract}
\section{Introduction}
Open Data from public administrations, also called \textit{Open Government Data} (OGD), provides a rich source of structured data that has become a key component of an evolving Web of Data. The key factors for the success of OGD initiatives are on the one hand the incentives for publishing organizations to demonstrate transparency or compliance to regulations, but on the other hand also the availability of agreed standards and best practices for publishing OGD:
de facto publishing standards for metadata on OGD portals such as \textit{DCAT} or, more recently, Schema.org's dataset vocabulary~\cite{schema_org_ref}, as well as widely used open publishing software frameworks such as \textit{CKAN} or \textit{Socrata}, provide technical means to publish structured data along with descriptive metadata. There are over 250 (governmental) portals worldwide relying on these software frameworks for describing and publishing datasets~\cite{DBLP:journals/jdiq/NeumaierUP16}.  Yet, as more and more data is becoming available, findability, as well as quality and trust are of utmost importance in order to utilize the data. While in terms of findability, metadata about the temporal and geo-spatial scope of datasets are most relevant~\cite{KACPRZAK201937,neumaier2018search}, \emph{provenance} information has to be known to assess the trustworthiness of OGD. 
This is usually done in the form of giving a speaking label of the publishing body in the metadata. For instance, "European Commission" is mentioned as a publisher of 12,448 datasets on \url{https://data.europa.eu/}; an organization that can be uniquely referenced also in existing knowledge graphs (KGs) such as DBpedia (\url{dbr:/European_Commission})\footnote{URL prefixes such as \texttt{dbo:}, \texttt{dbp:}, \texttt{wdt:}, or \texttt{schema:} can be referenced in \url{prefix.cc}.} or Wikidata (\url{Q8880}), however, such links are not (yet) explicit. Moreover, in other cases, different publishing organizations within the metadata have non-descriptive names such as "Parlament"\footnote{German writing of the English word "parliament".} (on \url{https://data.gv.at/}), which only in the context of the portal itself make sense.\footnote{As \url{https://data.gv.at/} is the Austrian national data portal, the label "Parlament" refers to the Austrian parliament.}
Apparently, the publisher here actually refers to the \emph{Austrian Parliament}.
Alternatively, in other cases in addition, the contact information (e.g. an e-mail address or URL) found in the metadata can provide additional context on the publishing organization.

Summarizing, organizational information is usually not yet standardized in OGD portals by means of unique identifiers.
Notably, this problem is aggravated by the fact that public bodies -- just as any institution -- are affected by \emph{organizational changes}, that is, for instance ministries are being merged or restructured, and, therefore, the publishers may change over time and across different versions of datasets.
Overall, this means that, while several qualitative comparisons of Open Data initiatives exist on a country level\footnote{cf. for instance \url{http://opendatamonitor.eu} or\\  \url{http://europeandataportal.eu/dashboard}}, tracking the success of Open Data policies on the level of publishing organizations, or, respectively, tracking the development of these organizations in terms of mergers and re-structuring is hardly possible at the moment.

We argue that unambiguously linking Open Data publishers to URIs in public KGs would both increase findability of datasets (e.g. queries for datasets by statistical offices located in the European Union would be possible) as well as make it easier for data consumers to trust in the data, given reliable provenance information. Additionally, advanced queries for dataset monitoring and analyses would become possible. Lastly, even changes in organizations (such as mergers and renamings) would be less confusing for dataset users as long as the organizations still remain correctly linked. 
We therefore believe that linking Open Government Data and metadata with entities found in open KGs could be a solution to the stated ambiguity problems. Yet, to the best of our knowledge, neither the coverage of OGD publishing organizations, nor the specific challenges of this organizational linking problem have been investigated so far.
The focus of of the present position paper is therefore to study the feasibility and main open research problems for providing working solutions in this area. 
More concretely, we identify five challenges that are yet to be solved when linking organizations of public datasets to knowledge graphs, for each of which, we discuss potential solutions to be applied by Open Data publishers, the knowledge graph community, or -- where possible -- through automated linking approaches.

The rest of the paper is structured as follows: Section \ref{sec:RW} provides an overview of the most important related work to our target contributions. In Section~\ref{sec:performance} we analyze and briefly show the (non-)performance of state-of-the-art entity linking systems. This analysis is based on a gold standard that we created to further analyze and motivate underlying issues; subsequently, the identified main challenges in our opinion primarily responsible for this poor performance are discussed one by one in Section~\ref{sec:chal}, whereupon Section~\ref{sec:solutions} discusses possible directions and starting points to tackle them. We conclude in Section~\ref{sec:Conclusion} with an outlook and call for future work. 

\section{Background and Related Work}
\label{sec:RW}

Our analysis and observations are mainly based on the data gathered by the \textit{Open Data Portal Watch} (ODPW) project\footnote{\url{https://data.wu.ac.at/portalwatch}}. ODPW provides a large collection of Open Data metadata which has been compiled in order to monitor the quality of OGD portals: the ODPW project is regularly collecting metadata from over 250 portals world-wide, providing access to metadata dumps as weekly snapshots as well as various quality metrics~\cite{DBLP:journals/jdiq/NeumaierUP16,neumaierPhD2019} per portal, i.e., typically at country-level. However, a more fine-grained analysis of Open Data quality, as well as analysis of Open Data \emph{on the level of single publishing organizations} is not yet supported, for the reasons we will outline in the following sections. 

As for other related work on connecting Open Data to KGs, in~\cite{DBLP:conf/i-semantics/ErmilovAS13} the authors propose a system to integrate user-generated mappings of attributes into an existing Open Data ecosystem; however, this system did not yet allow links to public KGs. Moreover, the temporal aspects of changes, that we will focus upon herein, were not covered there. The system was part of the former EU Open Data Portal~\footnote{\url{https://data.europa.eu/euodp/de/data}} but is currently not available there anymore.

Tygel et al.~\cite{DBLP:conf/semco/TygelADOC16} present a system to link datasets from different Open Data portals by extracting the tags and keywords from metadata descriptions: the tags get reconciled using automated translations and similarity measures, and re-published using unique URIs and meta-information for the reconciled tags. Again, specifically, links to organizations and temporal changes were not taken into account in this approach. However, the approach tries to solve multilinguality-issues using automated machine translation, which as we will discuss below, are also relevant in our context.

Overall, we observe that so far not much work has been carried out in terms of matching organizational information from OGD datasets to KGs. The most prominent recent contribution is the Google Dataset Search~\cite{DBLP:conf/www/BrickleyBN19} service which offers a dedicated search engine for public datasets. To do so, Google links the identified datasets to their internal knowledge graph, in particular by partially mapping the publishing organizations. While no details about the actual matching approach and its coverage are provided, as a main challenge (besides data quality) they identify the ambiguity of organization names which are tackled by considering the website context for the mappings.

Related to our addressed challenge of linking organisations in the context of OGD is the heterogeneity of academic/research organisations; this is addressed by the EU-funded project RISIS.\footnote{\url{http://risis.eu/orgreg/}} 
The goal of this project is to provide a comprehensive register of public-sector research and higher education organizations in European countries. 
Each entry in the register provides a stable identifier and a set of characteristics of the entities, such as the website, country, and the entity type. While the register is a valuable source in the domain of research organisations, there is no coverage in the domain of OGD.

Notably, there are already existing standards and vocabularies which aim to solve the problem of heterogeneous metadata and missing links to publishing organizations. For instance, the \textit{Semantic Government Vocabulary} (SGoV)~\cite{DBLP:journals/ws/KremenN19}, the \textit{Data Catalog Vocabulary} (DCAT)~\cite{dcat}, etc. -- for publishing OGD. Yet, in practice, where these vocabularies are used (or mapped to, as in ~\cite{neumaierPhD2019}) the respective attributes to link to publishing organizations are rather linking to (ambiguous) string labels than to URIs.

\section{On the Performance of Current Entity Linking Systems}
\label{sec:performance}
In order to analyze the challenges of linking organizational information to knowledge graphs' entities in depth, three methods have been applied which are presented in the following.

\subsection{Analysis of the ODPW Database}
To assess the linking problem quantitatively, we focus on metadata from the Open Data Portal Watch data base~\cite{DBLP:journals/jdiq/NeumaierUP16}, accessible via a public API. The data base consists of weekly metadata crawls from 252 data portals starting in early 2016. As of November 2019, when we conducted this analysis, the data covered 2,552,114 individual datasets. Under scrutiny for this work are the organizational metadata details and changes of those over the observed time frame. All statistical figures concerning organizational metadata in public datasets given in this paper refer to this metadata corpus unless stated otherwise.

\subsection{Gold Standard for Change Analysis and Linker Evaluation}
\label{ssec:gold_standard}
From the corpus, we created links from randomly chosen ODPW datasets by manually assigning the publishing organizations in terms of their existing Wikidata and DBpedia entities. We linked $200$ distinct organizations of 174 distinct datasets. A match was only added to the gold standard when at least a link to Wikidata could be found. Each link was checked by at least two authors of this paper and only added if there was agreement concerning the link. The annotated instances in the final gold standard are from 57 different data portals and cover publishing organizations in different parts of the world. Notably, out of these 200, only 72.5\% of the organizations could be manually matched to DBpedia which suggests a lower organization coverage for public administration institutions compared to Wikidata.

The gold standard also covers organizational changes in the datasets in terms of updates on the \texttt{dcat:publisher} property:\footnote{The ODPW metadata already maps different schemata uniformly to DCAT, cf.~\cite{neum-etal-LDOW2017}.} for 26 datasets direct changes can be observed in terms of updated label for the \texttt{dcat:publisher}; that is, 26 out of the 200 linked instances potentially reflect an organizational change, some of which could indeed be mapped to different organizations (whereas others only indicate a refinement or correction). The gold standard is publicly available on GitHub\footnote{\url{https://github.com/YaserJaradeh/LinkingODPublishers/blob/master/GoldStandard.csv}} under the CC-BY license. It can also be used to evaluate linking systems on their ability to match organizational entities.

\subsection{Evaluation of Current Matching Systems}
\label{ssec:simple_linker}
To make the point of limited usability of currently available entity linkers ``off-the-shelf'', we evaluated multiple state-of-the-art linkers (on the \texttt{dcat:publisher} label information only) and also implemented a na\"{i}ve baseline entity linker based on \emph{term frequency - inverse document frequency} ($TF/iDF$) by comparing whole metadata descriptions with DBpedia and Wikidata abstracts. 
For each target entity, i.e. an entity of a public KG, a document is built consisting of its labels, alternative forms, as well as its \texttt{dbo:abstract} in the case of DBpedia, and \texttt{schema:description} in case of Wikidata.
The linker produces a one-to-many mapping by ranking $TF/iDF$ matches in decreasing order according to their similarity scores. To obtain a one-to-one mapping, only the top-1 match is considered for the evaluation. The linker is implemented in Python and is available on GitHub\footnote{\url{https://github.com/YaserJaradeh/LinkingODPublishers/tree/master/Scripts}}.

In order to give a quick overview on the performance of current entity linking systems, seven state-of-the-art systems as well as our introduced baseline linker
have been run on the datesets of our manually created gold standard. While linking to Wikidata achieves generally better results due to a higher concept coverage, the performance scores even for the best linking systems are clearly too low for a fully automated approach. 

\begin{table}[]
\centering
\caption{Performance of different linking systems for the two target KGs Wikidata and DBpedia. The best F1 scores of each KG are represented in bold print. Most systems evaluated here are tailored to one specific target graph. The symbol (-) indicates that the system does not work on the stated knowledge graph.}
\label{tab:linker-performance}
\resizebox{0.8\textwidth}{!}{%
\begin{tabular}{lc|c|c||c|c|c}
                           & \multicolumn{3}{c||}{\textbf{DBpedia}} & \multicolumn{3}{c}{\textbf{Wikidata}} \\ \cline{2-7} 
                           & \textbf{P} & \textbf{R} & \textbf{F1} & \textbf{P}  & \textbf{R} & \textbf{F1} \\ \hline
\multicolumn{1}{l|}{\textbf{Exact Matching}}    & 0.071           &  0.059          &  0.063           & 0.099            & 0.102           &  0.1           \\
\multicolumn{1}{l|}{\textbf{DBpedia Spotlight}~\cite{DBLP:conf/i-semantics/MendesJGB11}} & 0.214      & 0.223      & 0.217       & -           & -          & -           \\
\multicolumn{1}{l|}{\textbf{Text Razor}~\cite{textrazor}}        & -          & -          & -           & 0.214       & 0.206      & 0.207       \\
\multicolumn{1}{l|}{\textbf{EARL}~\cite{DBLP:conf/semweb/DubeyBCL18}}              & 0.204           & 0.2            & 0.201            & -           & -          & -           \\
\multicolumn{1}{l|}{\textbf{TagMe}~\cite{DBLP:conf/cikm/FerraginaS10}}             & 0.055      & 0.067      & 0.06       & -           & -          & -           \\
\multicolumn{1}{l|}{\textbf{Meaning Cloud}~\cite{meaning_cloud}}     & 0.105      & 0.104     & 0.103       & -           & -          & -           \\
\multicolumn{1}{l|}{\textbf{FALCON}~\cite{DBLP:conf/naacl/SakorMSSV0A19}}            & 0.266      & 0.254      & 0.258       & -           & -          & -           \\
\multicolumn{1}{l|}{\textbf{Open Tapioca}~\cite{DBLP:journals/corr/abs-1904-09131}}      & -          & -          & -           & 0.432       & 0.42       & 0.423       \\ 
\multicolumn{1}{l|}{\textbf{Simple TF-IDF Linker}}        & 0.39       & 0.373      & \textbf{0.378}       & 0.621       & 0.587      & \textbf{0.596}      
\end{tabular}%
}
\end{table}

This small experiment clearly demonstrates that relying solely on exact string matching techniques, or comparing abstracts with metadata descriptions achieves poor results as shown in Table~\ref{tab:linker-performance}. We argue that the problem is indeed not purely based in the non-suitability of matching techniques themselves but fundamentally related to open challenges brought by the nature of organizational data and their representation in KGs. 

Though a simple TF/iDF approach is used for the na\"{i}ve linker, it is still outperforming other baselines. One reason is that this simple linker only searches a part of the entire search space, namely the collection of organizations found in the knowledge graph. Other general purpose tools (e.g. DBpedia Spotlight) try first to find out what the entity type is and perform the actual linking afterwards. Therefore, in the case of the simple linker, the search space is more restricted and the disambiguation process is more accurate.

\section{Challenges}
\label{sec:chal}

In the following section we identify fundamental and in our opinion open challenges that complicate automated linking of organizations to KGs; each identified challenge will be illustrated and where possible quantitatively analyzed on our corpus or, respectively the KGs under consideration (DBpedia and Wikidata).

\subsection{Challenge 1: Temporal Changes in Organizations}

\subsubsection{Challenge Statement} 
Organizations change over time due to mergers, splits, or renamings. On the other hand, the metadata of open datasets also changes over time. Both types of changes complicate the automated linking process.

\subsubsection{Challenge Analysis}
While the actual change in organizational structures is not in itself problematic, there are many consequences which affect the linking process. For instance, information about a change might not be reflected in the dataset's metadata and/or in public KGs, or respectively be reflected asynchronously: there is latency in terms of both when/whether the change is updated in the metadata and the KGs. Furthermore, it is not clear how such temporal changes shall be reflected in public KGs (see Subsection \ref{ssec:missing_base_ontology}). 

In terms of temporal changes on the metadata level, we analyzed changes of individual dataset publishers per dataset and data portal over time.
In order to take into account the sheer size of our metadata corpus and to increase the performance of our analysis, a heuristic was applied: only datasets where the organization label of the first occurrence is different from the organization label in the last occurrence were considered in this statistical evaluation. In total, 109,280 organizational changes could be identified in this way. 

\begin{figure}
\centering
\includegraphics[scale=0.6]{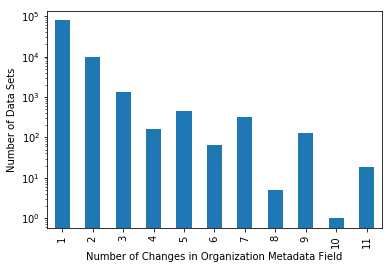}
\caption{The distribution of organizational changes for individual datasets. On the X-axis, the number of times an organization has been changed for a dataset on a particular data portal is shown while the overall frequency of such a change is reflected on the Y-axis. Note that the Y-axis is log-scale.}
\label{fig:distribution_of_changes}
\end{figure}

In a second step, the distribution of the number of organizational changes on a per dataset basis was analyzed. The maximum number of organizational changes is 11, meaning the organization of a single dataset on a data portal was changed 11 times. Figure~\ref{fig:distribution_of_changes} shows the distribution of changes. It can be seen that the distribution follows a power law: while 4 and more changes are relatively unlikely, datasets with one change clearly dominate the distribution with 78,378 occurrences.

\begin{figure}
\centering
\includegraphics[scale=0.6]{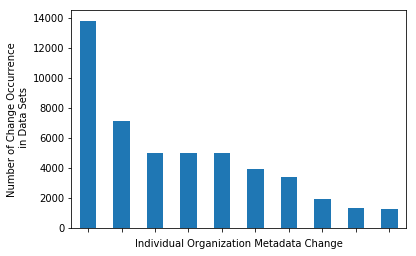}
\caption{The number of datasets in which an individual change, such as an organizational renaming, occurs. Each bar represents an individual change (cf. Table~\ref{tab:top_ten_changes}), e.g., the first one refers to the change from ``NSGIC GIS Inventory (aka Ramona)'' to ``NSGIC GIS Inventory'' and was propagated to almost 14000 individual datasets.}
\label{fig:distribution_of_org_changes}
\end{figure}

It was further analyzed how many organizational changes cause the changes on dataset level, i.e. whether there are bulk changes that propagate across different datasets and portals (e.g. through ``harvesting portals'' that import metadata from other portals), and how these changes are distributed. The fact that there are only 33,879 distinct organization labels in the dataset but that there are more than 90,000 changes on a per-dataset level, indicates that bulk changes occur. 
We found that these roughly 90,000 changes on dataset level are caused by 12,489 individual changes of organization labels. It is important to note that the number of changes here does not necessarily reflect changes on distinct organizations -- multiple renamings of the same organization are also counted (which can likewise be seen in Table~\ref{tab:top_ten_changes}).
Figure~\ref{fig:distribution_of_org_changes} shows the distribution of individual organization label changes. Although there are few changes that are heavily picked up in the datasets, the distribution is more linear compared to the one in Figure~\ref{fig:distribution_of_changes}. Table~\ref{tab:top_ten_changes} displays the ten most frequent label changes on ODPW. 

\noindent Overall, in our analysis of the Open Data Portal Watch data base, we qualitatively identify six reasons for organizational metadata changes -- only two of them (I and II) being an actual change of the organization: 

\begin{enumerate}[I]
\item \textbf{Renaming:} Organizational changes due to actual renamings of the publishing institution such as from \textit{Department for Communities and Local Government} to \textit{Ministry of Housing, Communities and Local Government}.
\item \textbf{Structural Changes:} Changes due to structural transitions of the publishing organization such as mergers, divisions, or other restructurings. An example here would be the \textit{Department of Energy and Climate Change} that was merged with the \textit{Department for Business, Innovation and Skills} to form the \textit{Department for Business, Energy and Industrial Strategy}. 
\item \textbf{Specialization:} Changes to further specify which organization is meant such as changing the label \textit{Department of Education} to \textit{Department Of Education (Northern Ireland)}. In this category also fall changes that further define which part of the organization was involved in the dataset creation or provision such as from \textit{Bristol City Council} to \textit{Bristol City Council - Sustainability Team}.
\item \textbf{Generalization:} Changes that generalize the authorship -- most likely in order to make the publisher easier to find and to identify. An example here would be a change from \textit{Martin Farrell} to \textit{West Sussex County Council}.
\item \textbf{Editorial/Error Correction:} Changes due to corrections e.g. from \textit{Ordance Survey} to \textit{Ordnance Survey}.
\item \textbf{Other:} Changes that we could not further classify, such as from \textit{Frederick Manby, Noah Linden} to \textit{Science} or from \textit{Daryl Beggs, Ruth Oulton, Benjamin Lang} to \textit{Benjamin Lang, Daryl Beggs, Ruth Oulton}, neither of which are mappable to an actual organization.
\end{enumerate}

\subsection{Challenge 2: Lack of Consistently used Base Ontology for Modeling Organizations in  Knowledge Graphs}
\label{ssec:missing_base_ontology}

\subsubsection{Challenge Statement} Organizations change over time -- however, this change cannot be sufficiently expressed in current knowledge graphs' ontologies or, respectively, existing capabilities of the vocabulary are not broadly used. Additionally, the vocabulary is insufficient in terms of expressing the relation between a geographic area and its governing body.

\subsubsection{Challenge Analysis} 
Wikidata offers specific properties for capturing different types of organizational changes. For example, \textit{Also Known As} (\texttt{skos:alt\-La\-bel}), \textit{Official Names} (\texttt{wdt:P1448}), \textit{Replaces} (\texttt{wdt:P1365}), \textit{Replaced By} (\texttt{wdt:P1366}), and \textit{Follows} (\texttt{wdt:P155}). \textit{Wikidata}'s model also offers a property named \textit{The Point In Time} (\texttt{wdt:P585}) which indicates the date from when a fact considered \texttt{true}. However, in the majority of the studied instances, these properties were not used to reflect organizational changes. In only 50\% of all cases, an organizational change was reflected in Wikidata -- mainly by using \textit{Also Known As} and \textit{Official Names} properties. For instance, the \textit{Department for Environment and Water} (\texttt{wdt:Q5260295}) formally known as \textit{Department of Environment, Water and Natural Resources}, is correctly listed in Wikidata using the property \textit{official names} along with the \textit{End Time} property (\texttt{wdt:P582}). However, other captured changes are only annotated as \textit{Also Known As} without any further details. A case in point is \textit{Department of Finance, Services and Innovation} (\texttt{wdt:Q17004340}) previously known as \textit{Department of Finance and Services}. While on the level of countries and states, the relation to governing or administrative bodies is clear through multiple concepts used with multiple properties (e.g. \textit{office held by head of state} (\texttt{wdt:P1906}) or \textit{applies to jurisdiction} (\texttt{wdt:P1001})), this is not true anymore for smaller areas such as cities. Here, \textit{applies to jurisdiction} (\texttt{wdt:P1001}) is typically used. In many cases, there is no distinction made between the area and the governing body.

DBpedia's vocabulary also offers properties to capture organizational changes, such as \texttt{dbp:preceding}, \texttt{dbp:replace}, \texttt{dbp:predecessor},
\texttt{dbp:suc\-ces\-sor}, and property \texttt{dbp:merger}. However, such properties are not used widely to reflect actual changes. Furthermore, DBpedia lacks the temporal dimension that can be easily expressed in Wikidata. For instance, there is a property \texttt{dbo:mergerDate} -- however, this property's \texttt{rdfs:domain} is \texttt{dbo:Place} which is, in fact, disjoint with \texttt{dbo:Or\-ga\-ni\-sa\-tion}'s parent class \texttt{dbo:Agent} and consequently cannot be used to express the temporal details of an organizational merger. To quantify these statements, in the sampled  organizational changes, only 34\% of the conducted organizational changes were reflected in DBpedia entities. Moreover, similar to Wikidata, changes are rarely supported with details that describe the change. For instance, \textit{London Fire and Emergency Planning Authority}\footnote{\texttt{dbr:London\_Fire\_and\-\_Emergency\-\_Planning\_Authority}} 
which replaces \textit{London Fire and Civil Defence Authority}\footnote{\texttt{dbr:London\_Fire\_and\_Civil\_Defence\_Authority}}, 
is captured by \texttt{dbp:predecessor} and \texttt{dbp:successor} properties in both instances -- however, without any timeline information. In terms of the relation between geospatial areas and governing bodies on DBpedia, it is available to some extent in the vocabulary through multiple properties such as \texttt{dbp:governingBody}, but the vocabulary is rarely used.\footnote{A SPARQL query for \texttt{dbp:governingBody} resulted in $\sim6,000$ usages with only 930 distinct objects over all of DBpedia.} Similar to Wikidata, in many cases no distinction is made between areas and the governing body. We note even that the DBpedia entities themselves are inconsistent in this regard, such as \texttt{European\_Union} both being typed as \texttt{dbo:Country}
and \texttt{dbo:Organisation}, two classes labelled as \texttt{owl:disjoint\-With}.

\subsection{Challenge 3: Metadata Quality}

\subsubsection{Challenge Statement} Varying metadata quality among data portals complicates the automated linking process.

\subsubsection{Challenge Analysis} 
Poor metadata quality in provenance information is a major issue when linking organizations to unique KG entities. This issue has also been addressed by Google~\cite{DBLP:conf/www/BrickleyBN19} and could be confirmed by our analysis of the ODPW data base: as an example, Table \ref{tab:top_ten_changes} displays the most frequent organization metadata changes on ODPW. It can be noted that from the 10 most frequent changes, at least 4 are not meaningful, and in fact loosing semantics (marked in \emph{italic}, i.e. not indicating any derferenceable publishing organization, but rather generic departments names or individual authors (potentially raising additional privacy problems).

\begin{table}[]
\centering
\caption{Top 10 organization changes by label together with the number of occurrences of the change within datasets listed in ODPW.}
\selectlanguage{russian} 
\resizebox{\columnwidth}{!}{%
\begin{tabular}{|l|l|c|}
\hline
\textbf{Old Organization Label}                     & \textbf{New Organization Label}                     & \textbf{Frequency} \\ \hline
NSGIC GIS Inventory (aka Ramona)                    & NSGIC GIS Inventory                                 & 13,793              \\ \hline
Geoscience Australia                                & \emph{$Corp$}                                                & 7,111               \\ \hline
Daryl Beggs, Ruth Oulton, Benjamin Lang,            & \emph{$Benjamin$ $Lang$, $Daryl$ $Beggs$, $Ruth$ $Oulton$,}            & 5,007               \\ \hline
Daryl Beggs, Ruth Oulton, Benjamin Lang,            & \emph{$Engineering$}                                         & 5,007               \\ \hline
Benjamin Lang, Daryl Beggs, Ruth Oulton,            & \emph{$Engineering$}                                         & 5,007               \\ \hline
Ivan Begtin                                         & Федеральная служба статистики                       & 3,359               \\ \hline
Archive bot                                         & Национальный цифровой архив России                  & 1,925               \\ \hline
Senatsverwaltung für Gesundheit und Soziales        & Senatsverwaltung für Gesundheit und Soziales Berlin & 1,298               \\ \hline
Senatsverwaltung für Gesundheit und Soziales Berlin & Senatsverwaltung für Gesundheit und Soziales        & 1,273               \\ \hline
PAT S. Statistica & ISPAT & 1,121 \\ \hline

\end{tabular}%
}
\selectlanguage{english} 
\label{tab:top_ten_changes}
\end{table}

\subsection{Challenge 4: Multilinguality}
\label{ssec:c5_multilinguality}

\subsubsection{Challenge Statement}
\label{ssec:multilinguality_statement} As public dataset providers are spread around the world, different language identifiers further complicate the linking process. For example, the Chinese Central Bank is called \textit{People’s Bank of China} in English, \textit{Chinesische Volksbank} in German, and \begin{CJK*}{UTF8}{gbsn}中国人民银行\end{CJK*} in Chinese.

\subsubsection{Challenge Analysis}
The analysis of the ODPW data base showed that organizational labels are typically stated in the language where the publishing institution resides and that translations are often not given. An exception here is the European Data Portal (\url{europeandataportal.eu}): this portal harvests datasets from all member states, and the labels are automatically translated to English. However, automated translations do not necessarily correspond to the correct labels in other languages. DBpedia is not entirely multilingual in a sense that multiple labels are given in various languages for organizations in all cases: instead, there are dedicated DBpedia versions for multiple languages.  
Wikidata is more aligned in this regard: it is possible to define multiple labels in any given language. The People's Bank of China, for instance, can also be found using its Chinese label.
Even though multilingual labels can be defined on Wikidata, this is often not sufficient for our case. 
\selectlanguage{russian} 
For example, the concept \emph{Russian Federal State Statistics Service}
(Федеральная служба статистики) \emph{does} exist on Wikidata\footnote{\url{https://web.archive.org/web/20190403150124/https://www.wikidata.org/wiki/Q2624680}} -- but there is no Russian label defined for it as of April 2020. Yet, this label appears more than 3,000 times in the ODPW data base as publisher.
\selectlanguage{english}

\begin{figure}
\centering
\includegraphics[scale=0.6]{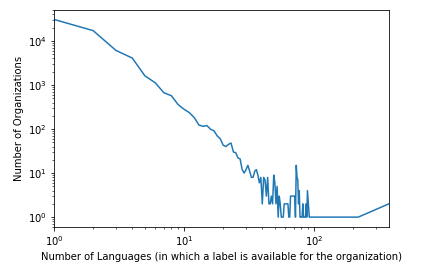}
\caption{Distribution of the number of languages in which labels exist for Wikidata concepts of type \textit{Organization}. The axes are logarithmically scaled. The distribution follows a power-law.}
\label{fig:wikidata_language_distribution}
\end{figure}

Figure~\ref{fig:wikidata_language_distribution} shows the distribution of languages in which a label is given for Wikidata entities typed as organization on a logarithmic scale. As it can be seen, the distribution follows a power-law: For most organizations labels are defined only in a single language. 

\subsection{Challenge 5: Disambiguating Public Sector Organizations}

\subsubsection{Challenge Statement} While companies can mostly be linked to one named entity without too much effort, this task is harder for public bodies. 

\subsubsection{Challenge Analysis} The disambiguation problem is two-sided: 
(i) When only states and cities are quoted as originator of a dataset, there is ambiguity in terms of the actual concept that is referred to in the KG which may hold multiple entities for a particular label. For example: Does \textit{New York} refer to the city of New York, the state of New York, or some particular administrative body of New York City? Wikidata contains entities for all three cases but the disambiguation is complicated without further context.
(ii) Similarly, given a concept in a KG, it can be hard to link it to dataset publishers due to the ambiguities -- in particular when acronyms are uses; this has also been pointed out by Google \cite{DBLP:conf/www/BrickleyBN19}. Also institution names common in several countries (e.g. ``\emph{Statistics Office}'' could be hard to disambiguate, though the portal context of nationally operated OGD portals may help here.\\

\section{Towards Solutions for Linking Organizations}
\label{sec:solutions}
In order to tackle the aforementioned challenges we outline possible solutions paths and starting points below.

\subsection{Challenge 1: Temporal Changes in Organizations}
\label{ssec:s1_temporal_changes}
The fact that organizations change over time is a given. It is, therefore, important to acknowledge this and improve current technologies, including publicly available KGs and data portals, to better and more consistently represent such changes. For example, on data portals, the metadata should be timestamped so that it is clear as of which date the information is valid. It may, in addition, help not to only store the most recent version but to keep a history of organizational labels and metadata since a change may not yet be reflected in the target to which the organization is mapped. 

\subsection{Challenge 2: The Lack of a Base Ontology for Public Knowledge Graphs}
\label{ssec:s2_missing_base_ontology}
At least as important as the design of a capable base ontology is the application. Our analysis showed, for instance, that the existing capabilities of the Wikidata and DBpedia vocabulary to reflect organizational changes are rarely exploited. 
Therefore, in order to automate and maintain mappings of such information, efforts to semantically represent organizational changes within KGs are required. This can be done by promoting currently available properties that express changes in organizations e.g. in the form of best practices for editors.
At the same time, existing ontologies can be extended to better capture organizations and administrative units.

\subsection{Challenge 3: Metadata Quality}
\label{ssec:s4_metadata_quality}
Improving the quality of metadata across different open datasets will significantly improve the linking quality. As mentioned earlier, while some of the observed changes are meaningful, such as ``NSGIC GIS Inventory (aka Ramona)'' to ``NSGIC GIS Inventory'' or ``Ivan Begtin'' to \selectlanguage{russian} ``\textit{Федеральная служба статистики}''
(\emph{Russian Federal State Statistics Service}), there is a concerning number of random changes occurring across datasets such as ``Corp''. One way to improve the metadata that is provided is to refine information extraction methods used to create such datasets. Further\-more, dataset publishers should be urged and motivated to keep their metadata as current and as accurate as possible. 
Moreover, as we are aware that the data quality will not improve instantly, automated linking systems need to be able to handle a certain amount of noise.
As our analysis showed, the metadata changes rather frequently (this is also observed by Google~\cite{DBLP:conf/www/BrickleyBN19}) -- hence, it is important to monitor meaningful changes on regular basis. Dataset harvesters, such as ODPW, can be bene\-ficial for this purpose by regularly retrieving metadata and detecting potential changes.\footnote{Note that to a certain extend, up-to-date metadata is available e.g. through the ODPW data base that was also used for our analysis: \url{https://data.wu.ac.at/portalwatch/data}.}
\selectlanguage{english}

\subsection{Challenge 4: Multilinguality}
\label{ssec:s5_multilinguality}
Multilinguality can be addressed by exploiting the multilingual capabilities of the Semantic Web. The problem is less  pronounced for Wikidata compared to DBpedia.  An active promotion of multilingual content (in KGs as well as on knowledge portals) can help in overcoming multilingual issues. For example, if multiple organization labels in multiple languages for the same organization would be available in the dataset metadata as well as in KGs, an overlap which might lead to a match becomes more likely.
In addition, the use of dictionaries, such as \textit{WordNet}~\cite{fellbaum_wordnet} or \textit{Wiktionary}, may help in some cases. For DBpedia, interlanguage links could be exploited to allow for multilinguality to a certain extent. 

\subsection{Challenge 5: Disambiguating Public Sector Organizations}
\label{ssec:s6_disambiguating_public_orgs}
One of the most pronounced problems is the disambiguation of labels. Here, the context has to be very broad to also include, for instance, the local top level domain (indicator for the country), contact e-mail addresses, and data portal URLs. Specialized linking systems are required for this task as current generic solutions fail to successfully disambiguate organizations. We believe that exploring the context of the dataset during the linking process can provide more accurate predication of the linked resource. Information such as the portal ID or URL provides more indications of the organization's context such as the country. For example, utilizing the country of an organization which is often found as part of portal ID can improve linking accuracy: if two organizations have the same name but are coming from different countries (e.g. \textit{Ministry of Education}) the broader context helps to disambiguate them.

\subsection{Across Challenges 1-5: Enabling a Community-Driven Linking Process}
As outlined above, a fully-automated linking process is as of now not available. Therefore, we argue that while the community is working on improving the boundary conditions, a manual lookup service is required that allows a data science community as well as dataset publishers to annotate organizational links on a dataset level together with a community effort. This can be applied by allowing humans to use a voting function to ensure a high linking quality. We believe that such a service will not only improve the linking quality but will also help in maintaining the linking results overtime. This service might look similar to \url{www.prefix.cc}, where publishers can be quickly looked up given a dataset URL or a data portal URL together with a label. As the same unique labels are used on many portals, the service could transitively reason organizational links for datasets not yet annotated which can be up or down voted by the user community. Over time, a larger gold standard could be created to improve and fine-tune existing linking systems. 

\section{Conclusion and Outlook}
\label{sec:Conclusion}
In this paper, we discussed the need and open problems of linking
organizations of public datasets to their corresponding entities in public KGs such as Wikidata and DBpedia. In order to understand the current state of this issue, we created a gold standard mapping of open dataset organizations to KGs entities. We evaluated the performance of different current entity linking approaches including our own simple approach. As the results of the automated linking approaches were disappointing,
we outlined five major challenges to be addressed. This includes (1) the temporal changes that happen on a regular basis in the organizations themselves and therefore in the metadata of open datasets. Our analysis also shows (2) that KGs are not fully using their existing capabilities to express organizational changes and we also address shortfalls in the existing vocabulary. We have also addressed (3) metadata quality issues and (4) multilinguality aspects within the linking process. Lastly, we found that (5) the disambiguation of public sector organizations is a hard task. 
We provide directions in terms of how these challenges can be addressed in the future. For future work, we aim to explore the idea of a community-driven effort in order to improve linking quality and to maintain that linking over time.

\subsubsection{Acknowledgements.} The authors thank Vincent Emonet, Paola Espinoza-Arias, and Bilal Koteich who contributed preliminary analyses regarding the challenges addressed in this paper. We also thank the organizers of the International Semantic Web Summer school (ISWS) 2019: the idea for this paper origins in discussions at the school.
\bibliographystyle{splncs04}
\bibliography{references}

\end{document}